\documentclass[aps,preprint,eqsecnum,showpacs]{revtex4}
\usepackage{amsfonts}
\usepackage{amsmath}
\usepackage[latin1]{inputenc}
\usepackage[T1]{fontenc}
\usepackage{textcomp}
\usepackage{times}
\usepackage[dvips]{color}

\begin{document}
\title{
Finite-size scaling in anisotropic systems }

\author{N.S. Tonchev}
\email{tonchev@issp.bas.bg}
\affiliation{Institute of Solid State Physics, 72 Tzarigradsko Chauss\'ee,
1784 Sofia, Bulgaria}
\begin{abstract}
We present analytical results for the  finite-size scaling in
$d$--dimensional $O(N)$ systems with strong anisotropy  where the
critical exponents  (e.g. $\nu_{||}$ and $\nu_{\perp}$) depend on
the direction. Prominent examples are systems with long-range
interactions, decaying with the interparticle distance $r$ as
$r^{-d-\sigma}$ with different exponents $\sigma$ in corresponding
spatial directions, systems with space-"time"a anisotropy near a
quantum critical point and systems with Lifshitz points. The
anisotropic properties involve also the geometry of the systems. We
consider systems confined to a d-dimensional layer with geometry
$L^{m}\times\infty^{n}; m+n=d$ and periodic boundary conditions
across the finite $m$ dimensions. The arising difficulties are
avoided using a technics of calculations based on the analytical
properties of the generalized Mittag-Leffler functions.
\end{abstract}
\maketitle

\section{Introduction}
Anisotropic systems are omnipresent in soft matter and solid state
physics. Prominent examples are liquid crystals, dipolar-coupled
uniaxial  ferromagnets, systems with Lifshitz points, systems with
space -"time" anisotropy near a quantum critical point, systems with
long-range interactions decaying with the interparticle distance $r$
as $r^{-d-\sigma}$ with different exponents $\sigma$ in
corresponding spatial directions and some dynamical systems
\cite{JC,MH02}. Specific problems arise in the consideration of
critical phenomena in such systems. In any of the cases the
fundamental idea of scaling must to be modified in a appropriate
way. In terms of the correlation length one can distinguish two type
of anisotropy: weak and strong. In weakly anisotropic systems, the
correlation length has spatially dependent amplitude. In the
strongly anisotropic systems, in addition the critical exponents
(e.g. $\nu_{||}$ and $\nu_{\perp}$) depend on the direction.  A more
general definition based on an anisotropic scale covariance of the
n-point correlators and different exhaustive examples one can see in
ref \cite{MH02}. In \cite{MH02} a general approach to scale
invariance in infinite volume systems with strong anisotropy has
been developed.

The object  of the present paper is scaling in finite - size
systems. In contrast to the theory of finite-size scaling in
isotropic systems (see e.g. \cite{brankov2000,chamati2003}) and
weakly anisotropic systems (see e.g \cite{CD04} and refs. therein)
the theory of finite - size scaling in strongly anisotropic systems
(see \cite{CT2000,CDT,BW89,L91,L92,HS01,H02,CGGP} ) is still a field
where the lack of results obtained in the framework of simplified
and analytically tractable models are noticeable. There exist by now
quite a few  examples  \cite{CT2000,CDT,CGGP} where the predictions
of anisotropic finite-size scaling hypothesis have been reproduced
analytically. In \cite{CT2000,CDT} anisotropy appears near a quantum
critical point as a result of mapping of a "time" dependent problem
(in d dimensions) to a "static" problem (in $d+1$ dimensions). In
\cite{CGGP} it is due to the spatial direction dependence of the
interactions .

Recently \cite{ChT05} a recipe for studying finite-size effects
based on some useful  properties of the generalized Mittag-Leffler
functions is suggested. It permits to consider isotropic and some
strongly anisotropic systems (including long-range quantum systems)
on an equal footing. The interest in Mittag-Leffler functions has
grown up because of their applications in some finite-size scaling
problems (see e.g.
\cite{brankov89,chamati1994,brankov2000,chamati2003,ChT05}). The
present study (see also \cite{T05}) is an illustration of the rare
possibility to handle the
 final expressions of the  scaling equations for strongly
 anisotropic systems analytically.

\section{The model}
We restrict our attention to the N-vector spin model defined at the
sites of the lattice. The Hamiltonian of the model reads
\begin{equation}
H=-N\sum_{x,y}J(x-y)\overrightarrow{\sigma_{x}}.\overrightarrow{\sigma_{y}},
\end{equation}
where $\overrightarrow{\sigma_{x}}$ is a classical N-component unit vector
 defined at site $x$
of the lattice and the spin-spin coupling decays with different
power-laws in different lattice directions,
 We assume a d-dimensional
system with \textit{mixed} geometry $L^{m}\times\infty^{n}$ under
periodic boundary conditions in the finite dimensions. The
interaction between spins enters the expressions of the theory only
through its Fourier transform. We will consider the following
anisotropic small ${\bf q}$ expansion of the Fourier transform of
the spin-spin coupling:
\begin{equation}\label{ai}
J({\bf q})\simeq J(0)+a_{||}|{\bf q}_{||}|^{2\sigma}+a_{\perp}|{\bf
q}_{\perp}|^{2\rho},
\end{equation}
where the first $n$ directions (called ``parallel'' and denoted by
the subscript $||$) are extended to infinity and the remaining $m$
directions (called ``transverse'' and denoted by $\perp$) are kept
finite, with $m+n=d$ and $a_{\perp}$ and $a_{||}$ are metric factors
and $\rho,\sigma>0$. In finite directions the corresponding
summations are over the vector ${\bf q}_{\perp}=\{q_{\perp
1},...,q_{\perp m}\}$ that takes values in $\Lambda^{m}$ defined by
$q_{\perp \nu}=2\pi n_{\nu}/(aN_{0})$ and $-(N_{0}-1)/2\leq
n_{\nu}\leq (N_{0}-1)/2, \nu=1, ...,m$. In infinite directions the
sums are substituted with normalized integrals over the
corresponding part of the first Brillouin zone
$[-\frac{\pi}{a},\frac{\pi}{a}]^{n}$. For our further purposes let
us recall that the  finite linear dimension $L=N_{0}a$, in the
continuous limit,   means that the lattice spacing $a \rightarrow0$
and simultaneously $N_{0}\rightarrow\infty$. In our analysis we
accept $a_{\perp}=a_{||}=-1/2$. Finite-size scaling behavior of such
type of systems are considered in reference \cite{CGGP} (with
$0<\rho,\sigma<1$). In the large-N limit, the theory is solved in
term of the gap equation for the parameter $\lambda_{V}$ related
with the finite-volume correlation length of the system. The bulk
system is characterized by a vanishing $\lambda_{\infty}$, so that
the appropriately scaled inverse critical temperature
\begin{equation}\label{ngit}
\beta_{c}=\frac{1}{(2\pi)^{d}}\int_{[-\pi/a]^{d}}^{[\pi/a]^{d}}\frac{d{\bf
q}}{|{\bf q}_{\perp}|^{2\rho}+|{\bf q}_{||}|^{2\sigma}}
\end{equation}
is finite whenever the effective dimensionality $D=m/\rho +n/\sigma$
is greater than 2. The corresponding critical exponents  $\nu_{||}$
and $\nu_{\perp}$ associated with the behavior of the infinite
system are
\begin{equation}\label{nu}
 \nu_{||}=\frac{1}{\sigma(D-2)},\quad
 \nu_{\perp}=\frac{1}{\rho(D-2)}.
\end{equation}
For more details see ref.\cite{CGGP}.

\section{The gap equation for the reference system}
 For the system with mixed geometry $\infty^{n}\times L^{m}$ the gap equation
 has the form:
\begin{equation}\label{ng}
\beta=\frac{1}{(2\pi)^{n}}\frac{1}{L^{m}}\int_{[-\pi/a]^{n}}^{[\pi/a]^{n}}\sum_{{\bf
q}_{\perp}\in\Lambda^{m}}\frac{d^{n}{\bf q}_{||}}{|{\bf
q}_{\perp}|^{2\rho}+|{\bf q}_{||}|^{2\sigma} + \lambda_{V}}.
\end{equation}
Our analysis will be limited to system below the upper critical
dimension $D_{u}=4$ and lower critical one $D_{l}=2$.

From the physical point of view, {\it the infinite n-dimensional
system}, which has a finite size $L$ in the remaining $m$
dimensions, can be found in three qualitatively different situations
depending on the value of $\frac{\sigma}{n}$: (i) If
$2<\frac{n}{\sigma}$, then the system is above  its lower critical
dimension $d_{l}=2\sigma$ and, therefore, it exhibits a true
critical behavior. A crossover from n-dimensional to $D$-dimensional
critical behavior takes place when $L\to \infty$. (ii) In the
borderline case of $n=2\sigma$, the system is at its lower critical
dimension and may have only a zero-temperature critical point. (iii)
When $\frac{n}{\sigma}<2$, the system is below its lower critical
dimension and a (D-dimensional) critical behavior appears only in
the thermodynamic limit $L\to \infty$.

 In the present study we assume that there is no phase
transition for finite $L$, hence $n<2\sigma$. For $n<2\sigma$ and
$\lambda_{V}\rightarrow 0$, due to the convergence of the integral
in (\ref{ng}) over $d^{n}{\bf q_{||}}$, one can extend the
integration over all $R^{n}$ in consistence with the  underlying
continuum field theory.

Further the corresponding n - dimensional integral  can be presented
as
\begin{equation}\label{sph}
\frac{1}{(2\pi)^{n}}\frac{S_{n}}{L^{m}}\int_{0}^{\infty}\sum_{{\bf
q}_{\perp}\in\Lambda^{m}}\frac{p^{n-1}dp}{|{\bf
q}_{\perp}|^{2\rho}+p^{2\sigma} + \lambda_{V}},
\end{equation}
where $S_{n}=2(\pi)^{n/2}/\Gamma(n/2)$ is the surface of the
$n$-dimensional unit sphere. With the help of the identity:
\begin{equation}\label{Scr2a}
\int_{0}^{\infty}\frac{p^{\alpha-1}dp}{t +p^{\eta}+|{\bf
q}_{\perp}|^{\tau}}=\frac{\Gamma(1-\frac{\alpha}
{\eta})\Gamma(\frac{\alpha}{\eta})}{\eta} \frac{1}{\left(t+|{\bf
q}_{\perp}|^\tau\right)^{1-\frac{\alpha}{\eta}}}, \qquad \eta >
\alpha>0,
\end{equation}
if we choose $\alpha=n$, $\tau=2\rho$ and $\eta=2\sigma$, for
Eq.(\ref{sph}) we end up with the result
\begin{equation}\label{564}
 \frac{A_{n,\sigma}}{L^{m}}\sum_{{\bf q}_{\perp}\in
\Lambda^{m}} \frac{1}{\left(\lambda_{V}+|{{\bf
q}_{\perp}}|^{2\rho}\right)^{1-\frac{n}{2\sigma}}}, \qquad 2\sigma
> n,
\end{equation}
where
\begin{equation}\label{k}
A_{n,\sigma}=\frac{S_{n}}{(2\pi)^{n}}\frac{\Gamma(1-\frac{n}
{2\sigma})\Gamma(\frac{n}{2\sigma})}{2\sigma}.
\end{equation}
Now the gap equation (\ref{ng}) may be presented in the equivalent
form
\begin{equation}\label{e11}
\beta= A_{n,\sigma}\frac{1}{L^{m}}\sum_{{\bf q}_{\perp}\in
\Lambda^{m}} \frac{1}{\left(\lambda_{V}+|{{\bf
q}_{\perp}}|^{2\rho}\right)^{1-\frac{n}{2\sigma}}}, \qquad 2\sigma
> n.
\end{equation}
Let us emphasize that one can relate Eq.(\ref{e11}) with a
fictitious {\it fully finite isotropic} $m$ dimensional reference
system in which the memory of the extended  to infinity dimensions
and of the anisotropy of the system is retained only in the
parameter
\begin{equation}\label{gam}
\gamma:={1-\frac{n}{2\sigma}},\qquad 0<\gamma<1,
\end{equation}
 and in
the multiplier $A_{n,\sigma}$ in front of the sum.

The normalized $m$ dimensional sum in Eq.(\ref{e11})
\begin{equation}\label{111}
W_{m,2\rho}^{1-\frac{n}{2\sigma}}(\lambda_{V},L):=\frac{1}{L^{m}}\sum_{{\bf
q}_{\perp}\in \Lambda^{m}} \frac{1}{\left(\lambda_{V}+|{{\bf
q}_{\perp}}|^{2\rho}\right)^{1-\frac{n}{2\sigma}}}, \qquad 2\sigma
> n.
\end{equation}
can be evaluated with the help of the identity \cite{ChT05}
\begin{equation}\label{identity}
\frac{1}{(\lambda_{V}+y^{\alpha})^{\gamma}}= \int_0^\infty
dte^{-yt}t^{\alpha\gamma-1}
E_{\alpha,\alpha\gamma}^\gamma(-\lambda_{V}t^\alpha ),
\end{equation}
in terms of  the generalized Mittag-Leffler function
$E^{\gamma}_{\alpha,{\gamma}\alpha}(z)$ (see appendix \ref{MLf}). If
one chooses $\alpha=\rho$, $\gamma={1-\frac{n}{2\sigma}}$ and
$y=|{{\bf q}_{\perp}}|^{2}$ the needed result is

\begin{equation}\label{HCNT}
W_{m,2\rho}^\gamma(\lambda_{V},L)=
\int_{0}^{\infty}dx x^{\gamma\rho -1}
E^{\gamma}_{\rho,\gamma\rho}(-\lambda_{V}x^{\rho})
\left[\frac{1}{L}\sum_{q\in\Lambda^{1}}\exp(-q^{2}x)\right]^{m},\qquad
\gamma>0.
\end{equation}
Now let us define
\begin{equation}\label{135}
Q_{N_{0}}(x):=\frac{1}{L}\sum_{q\in\Lambda^{1}}\exp(-q^{2}x)=
\frac{1}{aN_{0}}\sum^{{N_{0}/2-1}}_{l=-N_{0}/2}\exp\left(-\frac{4\pi^{2}
l^{2}x}{a^{2}N_{0}^{2}}\right)
\end{equation}
and using the approximating formula (5.5) of ref.\cite{BD91}, we
have the expression
\begin{equation}\label{dan}
Q_{N_{0}}(x)\cong\frac{1}{\sqrt{4\pi x}}\left[\textrm{erf}
\left(\frac{\pi x^{1/2}}{a}\right)\right]-\frac{2\pi^{2}x}{3}
\frac{1}{a}\exp\left[-\left(\frac{\pi}{a}\right)^{2}x\right]+
\frac{1}{\sqrt{\pi
x}}\left\{\sum_{l=1}^{\infty}\exp[-(laN_{0})^{2}/4x]\right\}
\end{equation}
valid in the large $N_{0}$ asymptotic regime. The first and the
second terms in the above equation are size independent and are
precisely the infinite volume limit of $Q_{N_{0}}(x)$. The remainder
of the calculation involves the insertion of (\ref{dan}) into
(\ref{HCNT}).

\section{Finite-size scaling form of the gap equation}
In order to illustrate the further calculations we will consider in
more details the case $m=1$.
 We can represent the right-hand side of Eq.
(\ref{HCNT}) as a sum of three terms.

The first one is given by
\begin{equation}
\frac{1}{2\pi }
\int_{-\frac{\pi}{a}}^{\frac{\pi}{a}}dk\int_{0}^{\infty}dx
x^{\gamma\rho -1}
E^{\gamma}_{\rho,\gamma\rho}(-\lambda_{V}x^{\rho})\exp(-xk^{2})=
\frac{1}{2\pi } \int_{-\frac{\pi}{a}}^{\frac{\pi}{a}}dk
\frac{1}{\left(\lambda_{V}+k^{2\rho}\right)^{1-\frac{n}{2\sigma}}},
\end{equation}
where the definition of the erf-function
\begin{equation}\label{erf}
\frac{\textrm{erf}(\Lambda\sqrt{x})}{\sqrt{4\pi
x}}=\frac{1}{2\pi}\int_{-\Lambda}^{\Lambda}\exp(-xk^{2})dk
\end{equation}
and the identity (\ref{identity}) have been used.

 The second term is
\begin{equation}
-\frac{2\pi^{2}}{3a} \int_{0}^{\infty}dx x^{\gamma\rho}
E^{\gamma}_{\rho,\gamma\rho}(-\lambda_{V}x^{\rho})
\exp\left[-\left(\frac{\pi}{a}\right)^{2}x\right]=
-\frac{2\gamma\rho}{3}\frac{(\frac{\pi}{a})^{2\rho-1}}{[\lambda_{V}+
(\frac{\pi}{a})^{2\rho}]^{\gamma+1}}.
\end{equation}

The third one equals
\begin{equation}\label{thu}
\int_{0}^{\infty}dx x^{\gamma\rho -1}
E^{\gamma}_{\rho,\gamma\rho}(-\lambda_{V}x^{\rho})\frac{1}{\sqrt{\pi
x}}\left\{\sum_{j=1}^{\infty}\exp[-(jN_{0}a)^{2}/4x]\right\}.
\end{equation}
The first term is exactly the bulk limit
$W_{1,2\rho}^\gamma(\lambda_{V},\infty)$. The second one in the
considered regime $\lambda_{V}\rightarrow 0$ and $a\rightarrow 0$ is
of order $O\left(\frac{1}{\pi/a}\right)$ and can be omitted.

It is convenient  to write  the third term, Eq.(\ref{thu}), in terms
of the function (particular case of the Jacobi $\Theta_{3}$
function)
\begin{equation}\label{jad}
A(x)\equiv\sum_{n=-\infty}^{+\infty}e^{-xn^2}.
\end{equation}
and the universal finite-size scaling function \cite{F}
\begin{equation}\label{fssshift}
F_{m,2\rho}^\gamma(y)= \frac1{(2\pi)^{2\gamma\rho}} \int_0^\infty
dxx^{\gamma\rho-1} E_{\rho,\gamma\rho}^\gamma\left(-\frac
{x^{\rho}}{(2\pi)^{2\rho}} y\right) \left[A^{m}(x)-1-\left(\frac\pi
x\right)^{\frac {m}{2}}\right].
\end{equation}
This can be done with the help of the Poisson transformation formula
\begin{equation}\label{poiss}
A(x)=\sqrt{\frac{\pi}x}A\left(\frac{\pi^2}x\right)
\end{equation}
and the identity (\ref{inti}).

After some algebra the result for the third term is
\begin{equation}\label{etre}
L^{2\gamma\rho-1}\left[F^{\gamma}_{1,2\rho}(\lambda_{V}L^{2\rho})
+\frac{1}{(\lambda_{V}L^{2\rho})^{\gamma}}\right].
\end{equation}
Collecting the above results  for Eq.(\ref{HCNT}), if $m=1$,  we
obtain

\begin{eqnarray}\label{HCNT1a}
W_{1,2\rho}^\gamma(\lambda_{V},L)&=& \frac{1}{2\pi }
\int_{-\frac{\pi}{a}}^{\frac{\pi}{a}}dk
\frac{1}{\left(\lambda_{V}+k^{2\rho}\right)^{1-\frac{n}{2\sigma}}}-
\frac{2\gamma\rho}{3}\frac{(\frac{\pi}{a})^{2\rho-1}}{[\lambda_{V}+
(\frac{\pi}{a})^{2\rho}]^{\gamma+1}} \nonumber\\
&+&L^{2\gamma\rho-1}\left[F^{\gamma}_{1,2\rho}(\lambda_{V}L^{2\rho})
+\frac{1}{(\lambda_{V}L^{2\rho})^{\gamma}}\right],\quad
\gamma\equiv1-\frac{n}{2\sigma}>0.
\end{eqnarray}

 In the last term of Eq.(\ref{HCNT1}) apart from the
factor $L^{2\gamma\rho-1}$ the intrinsic scaling combination
\begin{equation}\label{scv}
y=\lambda_{V}L^{2\rho}=(L/\xi_{\perp,L})^{2\rho}
\end{equation}
emerges, where $\xi_{\perp,L}$ is the finite-size transverse
correlation length (see \cite{CGGP}).
The limitation $m=1$ is not principal. If $m>1$, in view of
Eq.(\ref{dan}), the product $[Q_{N_{0}}(x)]^{m}$ in Eq.(\ref{HCNT})
contains sums of terms of the form
\begin{equation}\label{sm1}
\left\{\frac{1}{\sqrt{4\pi x}}\right\}^{m}\left\{\left[\textrm{erf}
\left(\frac{\pi x^{1/2}}{a}\right)\right]\right\}^{m'}
\left\{\exp[-\sum_{i=1}^{m-m'}(l_{i}aN_{0})^{2}/4x]\right\}
\end{equation}
with $1\leq m'\leq m-1$ and $l_{i}\neq 0, i=1,...,m-m'$. In such
terms the error function $\textrm{erf} \left(\frac{\pi
x^{1/2}}{a}\right)$ can be replaced by unity, since the exponential
function in the right-side of Eq.(\ref{sm1}) cuts off the
contribution from values of $x^{1/2}\ll aN_{0}$. Note that all the
other terms that contain as a multiplier $\frac{2\pi^{2}x}{3}
\frac{1}{a}\exp\left[-\left(\frac{\pi}{a}\right)^{2}x\right]$ can be
estimated. They are of order $O\left(\frac{1}{\pi/a}\right)$ and
must be omitted in the considered continuum limit. As a result
instead of Eq.(\ref{HCNT1a}) we get
\begin{eqnarray}\label{HCNT1}
W_{m,2\rho}^\gamma(\lambda_{V},L)&\simeq& \frac{1}{(2\pi)^{m} }
\int_{[-\frac{\pi}{a}]^{m}}^{[\frac{\pi}{a}]^{m}}
\frac{d^{m}\bf{k}}{\left(\lambda_{V}+\bf|{k}|^{2\rho}\right)^{1-\frac{n}{2\sigma}}}\nonumber\\
&+&L^{2\gamma\rho-1}\left[F^{\gamma}_{m,2\rho}(\lambda_{V}L^{2\rho})
+\frac{1}{(\lambda_{V}L^{2\rho})^{\gamma}}\right],\quad
\gamma\equiv1-\frac{n}{2\sigma}>0.
\end{eqnarray}

 If we introduce the notations
\begin{equation}\label{k1}
K:=K(\sigma,n,m)\equiv A_{n,\sigma}^{-1}\beta
\end{equation}
Eq.(\ref{e11}) can be rewritten as
\begin{equation}\label{e111}
K - K_{\infty}^{c}=W_{m,2\rho}^\gamma(\lambda_{V},L) -
W_{m,2\rho}^\gamma(0,\infty),
\end{equation}
where
\begin{equation}\label{ct}
K_{\infty}^{c}:=K_{\infty}^{c}(\sigma,\rho,n,m)\equiv
W_{1,2\rho}^\gamma(0,\infty)=\frac{1}{(2\pi)^{m}
}\int_{[-\frac{\pi}{a}]^{m}}^{[\frac{\pi}{a}]^{m}}d^{m}{\bf k}
\frac{1}{\left(|{\bf k}| ^{2\rho}\right)^{\gamma}}
\end{equation}
is the inverse critical temperature (normalized with $A_{n,\sigma}$)
of the "isotropic" bulk system.

The first term in the right hand side of Eq.(\ref{HCNT1}) can be
presented in the form
\begin{eqnarray}\label{139}
 W_{m,2\rho}^\gamma(\lambda_{V},\infty)&=&\frac{1}{(2\pi )^m}
\int_{[-\frac{\pi}{a}]^{m}}^{[\frac{\pi}{a}]^{m}}d^{m}{\bf k}
\frac{1}{\left(\lambda_{V}+|{\bf
k}|^{2\rho}\right)^{\gamma}}\nonumber\\ &\simeq& K_{\infty}^{c}+
\frac{S_{m}}{2(2\pi)^{m} }
\lambda_{V}^{\frac{m}{2\rho}-\gamma}\int_{0}^{\infty}dx
\frac{x^{\gamma\rho}-\left
(1+x^{\rho}\right)^{\gamma}}{x^{\gamma\rho+1-m/2}\left
(1+x^{\rho}\right)^{\gamma}}
\end{eqnarray}
valid for $\xi_{\perp,L}\gg a$.  The integral over $x$ converges,
provided $m>2\gamma\rho>m-2\rho$, and
\begin{equation}\label{ints}
\int_{0}^{\infty}dx \frac{x^{\gamma\rho}-\left
(1+x^{\rho}\right)^{\gamma}}{x^{\gamma\rho+1-m/2}\left
(1+x^{\rho}\right)^{\gamma}}=
\frac{1}{\rho}\frac{\Gamma(\frac{m}{2\rho})}{\Gamma(1-\frac{n}{2\sigma})}
\Gamma\left(1-\frac{n}{2\sigma}-\frac{m}{2\rho}\right).
\end{equation}

By substitution Eq.(\ref{HCNT1}) into Eq.(\ref{e111}), taking into
account the small-argument expansion Eq.(\ref{139}), for the gap
equation (\ref{e11}) we obtain the scaling form:
\begin{equation}\label{ges}
x\approx-a(n,m;\rho,\sigma)y^{D/2-1} + F^{\gamma}_{m,2\rho}(y)
+\frac{1}{y^{\gamma}},\qquad 4>D>2,
\end{equation}
where
\begin{equation}
x=L^{2\rho(D/2-1)}(K-K_{c}),\qquad y=(L/\xi_{\perp,L})^{2\rho}
\end{equation}
and
\begin{equation}\label{a}
a(n,m;\rho,\sigma)=-\frac{1}{2\rho}\frac{S_{m}}{(2\pi)^{m}
}\frac{\Gamma(\frac{m}{2\rho})}{\Gamma(1-\frac{n}{2\sigma})}
\Gamma\left(1-\frac{n}{2\sigma}-\frac{m}{2\rho}\right).
\end{equation}
 Our model study confirm the
phenomenological assumption \cite{HS01} that the finite-size scaling
behavior in systems with mixed geometry $L^{m}\times \infty^{n}$ is
governed by the "perpendicular" correlation length only.
\section{Finite-Size corrections}

Given the gap equation in scaling form, we are now in a position to
explore the various finite-size corrections.
 Here, we look at the different regimes:  the finite-size scaling regime defined by the condition
$y \sim 1$, crossover to the thermodynamic
 critical behavior $y\gg1$, and  the regime $y \ll 1$. In this section for
the sake of simplicity  we will consider the important particular
case of slab geometry, $m=1$.
\subsection{$y\sim1$}
In order to consider the case of $y\sim1$ we will use a new
representation  for $F_{1,2\rho}^\gamma(y)$ (see appendix
\ref{appendix})

\begin{equation}\label{pip}
F_{1,2\rho}^\gamma(y)=F_{1,2\rho}^\gamma(0)+
a(n,1;\rho,\sigma)y^{\frac{1-2\gamma\rho}{2\rho}}+
2\sum_{l=1}^{\infty}\frac{(4\pi^{2}l^{2})^{\rho\gamma}-
[y+(4\pi^{2}l^{2})^{\rho}]^{\gamma}}{(4\pi^{2}l^{2})^{\rho\gamma}[y+
(4\pi^{2}l^{2})^{\rho}]^{\gamma}},\qquad 1>2\gamma\rho.
\end{equation}
and rewrite Eq.(\ref{ges}) in a form suitable for obtaining the
shift of the bulk critical temperature. Substituting Eq.(\ref{pip})
in Eq.(\ref{ges}) ($m=1$)   we obtain the gap equation in the form
\begin{eqnarray}\label{HCNT3}
x\simeq F_{1,2\rho}^\gamma(0)
+2\sum_{l=1}^{\infty}\frac{(4\pi^{2}l^{2})^{\rho\gamma}-
[y+(4\pi^{2}l^{2})^{\rho}]^{\gamma}}{(4\pi^{2}l^{2})^{\rho\gamma}[y+
(4\pi^{2}l^{2})^{\rho}]^{\gamma}} +\frac{1}{y^{\gamma}}.
\end{eqnarray}
Therefore, when $K\rightarrow K_{\infty}^{c}$, simultaneously with
$L \rightarrow \infty$, in the way prescribed by the equation
\begin{equation}
K=K_{\infty}^{c}+\frac{x}{L^{2\rho(D/2-1)}}
\end{equation}
with $x=O(1)$,  the leading-order asymptotic form of
$\lambda_{V}$ is given by
\begin{equation}
\lambda_{V}\simeq \frac{y(x)}{L^{2\rho}},
\end{equation}
where $y(x)$ is the positive solution of Eq.(\ref{HCNT3}). Hence, at
the critical point $x=0$, we obtain
\begin{equation}
\xi_{\perp,L}=A(n,\rho,\sigma)L
\end{equation}
where $A(n,\rho,\sigma)=1/[y(0)]^{1/2\rho}$ is an universal
amplitude.  In systems with mixed geometry the existence of the
universal amplitude of the (finite-size) correlation length
$\xi_{\perp,L}$ on the level of the phenomenological scaling has
been suggested in \cite{HS01}. Here this qualitative statement is
made quantitative being a model confirmation of the generalized
 in \cite{HS01} Privman-Fisher hypothesis (c.f. with Eq.(23) in \cite{HS01}).

  When $y \to \infty$, for $2<D<4$, we may approximate the sum in
  Eq.(\ref{HCNT3}) by an integral:
\begin{equation}\label{TH}
\sum_{l=1}^{\infty}\frac{(4\pi^{2}l^{2})^{\rho\gamma}-
[y+(4\pi^{2}l^{2})^{\rho}]^{\gamma}}{(4\pi^{2}l^{2})^{\rho\gamma}[y+
(4\pi^{2}l^{2})^{\rho}]^{\gamma}}\simeq\frac{1}{(2\pi
z)^{2\gamma\rho}}\int_{0}^{\infty}dx
\frac{(\frac{x}{z})^{2\gamma\rho}-\left
(1+(\frac{x}{z})^{2\rho}\right)^{\gamma}}{(\frac{x}{z})^{2\gamma\rho+1/2}\left
(1+(\frac{x}{z})^{2\rho}\right)^{\gamma}},\quad
z:=y^{1/2\rho}/(2\pi).
\end{equation}

With the use of Eqs.(\ref{ints}) and (\ref{TH}) one finds the
asymptotic solution of Eq.(\ref{HCNT3}), which to the leading order
in $x\gg1$ is
\begin{equation}
x\approx-a(n,1;\rho,\sigma)y^{D/2-1}
\end{equation}
i.e. it recovers exactly the familiar {\it bulk hight-temperature}
result.
\subsection{$y \gg 1$}
The finite-size correction to the bulk critical behavior can be
extracted from the asymptotic form of the functions
$F_{d,\sigma}^\gamma(y)$  at large argument $y \gg 1$ (see
\cite{ChT05, T05}). The result is

\begin{equation}\label{dzef}
F_{1,2\rho}^\gamma(y)\simeq-y^{-\gamma} +\left[2\gamma(2\pi)^{2\rho}
\zeta(-2\rho)\right]y^{-(1+\gamma)},
\end{equation}
where $\zeta(\rho)$ is Riemann's  Zeta function.

 Using Eq.(\ref{dzef}) for the gap equation (\ref{ng}) we
obtain:
\begin{equation}\label{dgr}
x\approx-a(n,1;\rho,\sigma)y^{D/2-1} +\left[2\gamma(2\pi)^{2\rho}
\zeta(-2\rho)\right]y^{-(1+\gamma)},\qquad y\gg1.
\end{equation}
As one can see the finite-size effects governed by the second term
in the  right hand side of Eq.(\ref{dzef}) vary as an algebraic
power of the variable $y$. Since $\zeta(-2\rho)=0$ for $\rho= k$,
$k$ - a natural number, there are not power low dependent finite
size corrections if $\rho=k$.
 The case $0<\rho<1$ corresponds to the long-range
interaction. For $\rho=1$ corresponding to the short-range
interaction the
 result for the universal finite-size scaling function is

\begin{equation}
F_{1,2}^\gamma(y)\simeq-y^{-\gamma} +\left[\frac
{1}{2^{\gamma}\Gamma(\gamma)}\right]
y^{-\frac{\gamma}{2}}e^{-\sqrt{y}}
\end{equation}
that reflects with exponential fall of finite-size corrections in
Eq.(\ref{dgr}). And as long as $\rho>1$ and if it is not an integer,
the power low corrections take place  in the case of so-called
subleading LR interaction \cite{BD91} but with strong anisotropy.

\subsection{$y\ll1$}
Whenever Eq.(\ref{HCNT3}) has a solution $y\ll 1$, use can be made
of the asymptotic expansion
\begin{equation}\label{ma}
\sum_{l=1}^{\infty}\frac{(4\pi^{2}l^{2})^{\rho\gamma}-
[y+(4\pi^{2}l^{2})^{\rho}]^{\gamma}}{(4\pi^{2}l^{2})^{\rho\gamma}[y+
(4\pi^{2}l^{2})^{\rho}]^{\gamma}}=-\frac{\gamma
}{(2\pi)^{2\gamma\rho+2\rho}}\zeta(2\gamma\rho+2\rho)y +O(y^{2}).
\end{equation}
In obtaining the first term in the right-hand side of Eq.(\ref{ma})
the fulfilling of the condition $2\gamma\rho+2\rho>1$ is used.
 Taking into account Eq.(\ref{ma}), in the limit $ |x-
F_{1,2\rho}^\gamma(0)|\ll1$ one finds
\begin{equation}
y\simeq \frac{1}{|x- F_{1,2\rho}^\gamma(0)|^{1/\gamma}}.
\end{equation}

Let us remind, that when the number of infinite dimensions is less
than the lower critical dimension, the singularities of the bulk
thermodynamic functions are rounded and no phase transition occurs
in the finite-size system. Nevertheless, one can define a
pseudocritical temperature, corresponding to the position of the
smeared singularities of the finite-size thermodynamic functions,
and study its shift with respect to the bulk value of the critical
temperature.
 In the case under
consideration the first term in the right hand side of Eq.(\ref{HCNT3}) is identified with the shift of the
finite-size pseudocritical
temperature. Actually, for the sake of convenience here we study the quantity $K$.
The corresponding result for the pseudocritical
$K_{L}^{c}$ is
\begin{equation}
K_{L}^{c} -
K^{c}_{\infty}=L^{-\lambda}F^\gamma_{1,2\rho}\left(0\right),
\end{equation}
i.e. the shift critical exponent is $\lambda=1/\nu_{\perp}$ in
accordance with standard finite-size scaling conjecture, see
\cite{brankov2000}. The coefficient $F_{d-n,\sigma}^\gamma(0)$ can
be evaluated analytically as well as numerically for different
values of the free parameters $d$, $\rho$ and $\gamma$ using the
method developed in reference \cite{chamati2000l}.


\section{Conclusions}
The statement, that finite-size scaling in our system with mixed
 geometry $L^{m}\times\infty^{n}$
 is the naturally expected one,
is contained in \cite{CGGP} where different  shape dependent scaling
limits have been studied. In the present study we are much more
interested in the {\it explicit form} of
 the scaling equation in different regimes. For this aim an
 appropriate technics of calculation is developed.
We show \textit{how} the mathematical difficulties that arise in the
considered anisotropic
 model with mixed geometry
can be avoided. First, using the identity (\ref{Scr2a}) the problem
is effectively reduced to the corresponding isotropic one related to
a fully finite reference system, Eq.(\ref{e11}). A further step
 is the recognition that with the help of the identity (\ref{identity})
  the appearance of $\gamma\neq
1$ in the summand of the gap equation (\ref{e11}) is not an obstacle
for an analytical treatment. Knowledge of the properties of the
generalized Mittag-Leffler function allows to carry out all
calculations
 analytically.
We show that though the system is strongly anisotropic, the corresponding
gap equation Eq.(\ref{ges}) for the intrinsic scaling variable $y=\lambda_{V}L^{2\rho}$
 has a form similar to the isotropic case with geometry
$L^{D}$(c.f. with Eq.(4.115) in \cite{brankov2000}). We stress that
the finite-size $L$  is scaled  by the perpendicular correlation
$\xi_{\perp}$, only. This verified the Privman- Fisher hypothesis
for strongly anisotropic systems formulated in \cite{HS01}.

Further we conclude that, the finite-size contributions to the
thermodynamic
 behavior decay algebraically as
a function of L only if $ 0<\rho\neq k$, where $k$ is a natural
number. In the  case $\rho=1$,  the finite-size contributions decay
exponentially as a function of $L$. The phenomenon that the
so-called subleading terms (in our terminology the term with
$\rho>1$) lead to dominant finite-size contributions, being
unimportant in the bulk limit, was first discussed in ref.
\cite{BD91}. This characteristic feature of the long-range
interactions reveals also in our consideration.

\begin{acknowledgments}
This work is supported by the Bulgarian Science Foundation under
Project F-1402.
\end{acknowledgments}

\appendix
\section{Generalized Mittag-Leffler functions}\label{MLf}
The  generalized Mittag-Leffler functions are defined by the power
series \cite{prabhakar1971} (see also
\cite{ChT05,saxena2004,kilbas2004})
\begin{equation}\label{gfunction}
E_{\alpha,\beta}^\gamma(z)=\sum_{k=0}^\infty\frac{(\gamma)_k}
{\Gamma(\alpha k+\beta)}\frac{z^k}{k!}, \ \ \ \ \ \alpha, \beta,
\gamma \in \mathbb{C}, \ \ \ \ {\mathrm Re}(\alpha)>0,
\end{equation}
 where
\begin{equation}
(\gamma)_0=1,\
(\gamma)_k=\gamma(\gamma+1)(\gamma+2)\cdots(\gamma+k-1)
=\frac{\Gamma(k+\gamma)}{\Gamma(\gamma)}, \ \ \ \ k=1,2,\cdots
\end{equation}
Let us formulate  some  needed properties  of the generalized
Mittag-Leffler functions.

It might be useful to note the relationship
\begin{equation}\label{pr4}
-\frac{\textrm{d}}{\textrm{d}z}E_{\alpha,1}^\gamma(-z^{\alpha})=
\gamma z^{\alpha-1}E_{\alpha,\beta}^\gamma(-z^{\alpha}),
\end{equation}
which follows from the power-series representation. In obtaining
Eq.(\ref{etre}) we have taken into account  the identity
\begin{equation}\label{inti}
\int_0^\infty dx x^{\gamma\rho-1} E^\gamma_{\rho,\gamma\rho}
\left(-x^{\rho}\right)=1, \ \ \ \  \rho>0
\end{equation}
that follows by integration of Eq.(\ref{pr4}) over $z$ from zero to
infinity. Next, by subtracting and adding $1/\Gamma(\alpha\gamma)$
to the function $E^\gamma_{\rho,\gamma\rho}$ we obtain

\begin{equation}\label{identity1}
 \int_0^\infty
dte^{-zt}t^{\alpha\gamma-1}
\left[E_{\alpha,\alpha\gamma}^\gamma(-t^\alpha
)-\frac{1}{\Gamma(\alpha\gamma)}\right]=
\frac{z^{\alpha\gamma}-(1+z^{\alpha})^{\gamma}}{(1+z^{\alpha})^{\gamma}z^{\alpha\gamma}}.
\end{equation}

\section{Derivation of equation (\ref{pip})}\label{appendix}
 First, we represent the integral in
Eq.(\ref{fssshift}) as a sum of three terms ($m=1$). The first term
is given by
 \begin{equation}\label{frt}
 (y^{-\frac{1}{\rho}})^{\gamma\rho}\int_0^\infty
dtt^{\rho\gamma-1} \left[E_{\rho,\rho\gamma}^\gamma(-t^\rho
)-\frac{1}{\Gamma(\rho\gamma)}\right]
\left[A\left(\frac{4\pi^{2}t}{y^{\frac{1}{\rho}}}\right) -1\right]
\equiv S_{\rho,\gamma}(y^{\frac{1}{\rho}}),
 \end{equation}
 the second term is
 \begin{equation}\label{sct}
-\frac{1}{2\sqrt{\pi}y^{\gamma-1/2\rho}} \int_0^\infty
dtt^{\rho\gamma-3/2} \left[E_{\rho,\rho\gamma}^\gamma(-t^\rho
)-\frac{1}{\Gamma(\rho\gamma)}\right]
\equiv-\frac{1}{y^{\gamma-1/2\rho}}C_{\gamma,\rho}
 \end{equation}
 and the third one equals the constant (provided $1>2\gamma\rho$)
 \begin{equation}\label{ml}
 \frac{1}{\Gamma(\rho\gamma)}\frac1{(2\pi)^{\gamma2\rho}}\int_0^\infty
dxx^{\gamma\rho-1}\left[A(x)-1-\left(\frac\pi x\right)^{\frac
{1}{2}}\right]\equiv F_{1,2\rho}^\gamma(0).
 \end{equation}\label{tht}
 Let us now calculate the function
 $S_{\rho,\gamma}(y^{\frac{1}{\rho}})$ and the constant
 $C_{\gamma,\rho}$. Making use of the identity (\ref{identity1}),
we represent  Eq.(\ref{frt}) as
\begin{equation}
S_{\rho,\gamma}(y^{\frac{1}{\rho}})=2\sum_{l=1}^{\infty}\frac{(4\pi^{2}l^{2})^{\rho\gamma}-
[y+(4\pi^{2}l^{2})^{\rho}]^{\gamma}}{(4\pi^{2}l^{2})^{\rho\gamma}[y+
(4\pi^{2}l^{2})^{\rho}]^{\gamma}}.
\end{equation}
To calculate $C_{\gamma,\rho}$, in Eq.(\ref{sct}) we first write
\begin{equation}
t^{-1/2}=\frac{1}{\pi^{1/2}}\int_{0}^{\infty}dx x^{-1/2}e^{-tx}
\end{equation}
then, by using the identity (\ref{identity1}) we take the integral
over $t$, and then
\begin{equation}
C_{\gamma,\rho}=\frac{1}{2\pi }\int_{0}^{\infty}dx
\frac{x^{\rho\gamma}-(1+x^{\rho})^{\gamma}}{(1+x^{\rho})^{\gamma}x^{\rho\gamma
+1/2}},
\end{equation}
i.e. $  C_{\gamma,\rho}=-a(n,1;\rho,\sigma)$. Collecting  the
results for  Eqs.(\ref{frt}), (\ref{sct}) and (\ref{ml}) for
(\ref{fssshift}) we get Eq.(\ref{pip}).

\end{document}